\documentclass{JHEP}
\usepackage{epsfig}
\usepackage{amssymb}

\newcommand{\be}{\begin{equation}}
\newcommand{\ee}{\end{equation}}
\newcommand{\bear}{\begin{eqnarray}}
\newcommand{\eear}{\end{eqnarray}}
\newcommand{\ba}{\begin{array}}
\newcommand{\ea}{\end{array}}
\newcommand{\bean}{\begin{eqnarray*}}
\newcommand{\eean}{\end{eqnarray*}}

\newcommand{\CL}{{\cal L}} \newcommand{\CE}{{\cal E}}
 
\newcommand{\CH}{{\cal H}} 
\newcommand{\CG}{{\cal G}} 
 
  \newcommand{\CO}{{\cal O}}

\def\det{\mathop{\rm det}}
\def\d{{\rm d}}

\preprint{ KIAS P04031\\
SNUST 041101\\
UOSTP 04104\\
{\tt hep-th/0411099}}
\title{Exactly Soluble Dynamics of $(p,q)$ String \\
Near Macroscopic Fundamental Strings
}
\author{Dongsu Bak ${}^{a}$, Soo-Jong Rey ${}^{b}$ \&
Ho-Ung Yee ${}^{c}$\\
~~~~~~~~~~~\\
${}^a$ Department of Physics, University of Seoul, Seoul {\rm KOREA}\\
${}^b$ School of Physics, Seoul National University, Seoul 151-747 {\rm KOREA}\\
${}^c$ School of Physics, Korea Institute for Advanced Study, Seoul 130-722 {\rm KOREA} \\
\vskip0.4cm \email{\tt dsbak@mach.uos.ac.kr, \hskip0.5cm
sjrey@phya.snu.ac.kr, \hskip0.5cm ho-ung.yee@kias.re.kr} }
\abstract{
We study dynamics of Type IIB bound-state of a Dirichlet string and
$n$ fundamental strings in the background of $N$ fundamental
strings. Because of supergravity potential, the bound-state string
is pulled to the background fundamental strings, whose motion is
described by open string rolling radion field. The string coupling
can be made controllably weak and, in the limit $1 \ll g^2_{\rm st}
n \ll g^2_{\rm st} N$, the bound-state energy involved is small
compared to the string scale. We thus propose rolling dynamics of
open string radion in this system as an exactly solvable analog for
rolling dynamics of open string tachyon in decaying D-brane. The
dynamics bears a novel feature that the worldsheet electric field
increases monotonically to the critical value as the bound-state
string falls into the background string. Close to the background
string, D string constituent inside the bound-state string decouples
from fundamental string constituents.}
\begin{document}
\section{Introduction}
Decay of unstable D-branes \cite{tachyon}, whose consequences led to
deep insight into the string theory \cite{tachyon2}, is triggered by
rolling the open string tachyon \cite{rolling}. While there has been
considerable development in recent years, the very fact that
intrinsically stringy dynamics is involved has deterred much
progress. For one thing, the tachyon mass is intrinsically of string
scale, rendering analysis based on low-energy Born-Infeld or
Yang-Mills effective field theory descriptions incomplete. Much
insight to the dynamics was gained \cite{rolling, rolling2} after
the full-fledged string theoretic description --- such as the
boundary-state for rolling tachyon --- was available. Even in this
approach, variety of issues are left unanswered, especially,
concerning final state of decaying D-branes \cite{rolling,
radiation}, and it would be highly desirable to devise a setup in
which essential physics of rolling tachyon are retained yet rolling
dynamics can be studied in a relatively simpler manner.

A situation in which dynamics involved is quite analogous to the
tachyon rolling is the formation of non-threshold bound-states
\cite{witten, narain}. Of particular interest are the bound-states
formed out of NS-branes (either 5-branes or fundamental strings) and
D-branes \cite{witten, LeeRey}. During the process, unlike other
non-threshold bound-states formed purely out of D-branes, no open
string tachyon field is involved. Rather, the constituents are
simply pulled into each other. We will refer to such dynamics as
{\sl rolling radion}, where the radion refers to (one of) the
transverse scalar fields measuring the relative separation between
binary constituents of the bound-state.

In this work, we shall demonstrate that there exists a dynamically
clean system in which the radion potential, which is the direct
analog of the tachyon potential, is controllably small so that the
energy would be released entirely into {\sl massless} open or closed
string excitations. The system involves fundamental (F) strings and
D strings \footnote{It is straightforward to extend this system to
various other S- or T-dual configurations. The resulting dynamics of
rolling radion would exhibit essentially the same behavior.}, and is
given by a probe $(n, 1)$ string (involving $n$ F-strings and one
D-string) moving in the background of $N$ F-strings. Dynamics of the
rolling radion is describable by (1+1)-dimensional U(1) Born-Infeld
theory in the superselection sector of $n$ unit of displacement
flux.

There are several highly nontrivial and remarkable features to the
system. First of all, supergravity solution describing $N$ F-strings
shows that the string coupling decreases monotonically toward the
F-strings so it can be made small everywhere so that string
perturbation theory is reliable. Second, NS-NS $B$-field increases
monotonically to the critical value toward the F-strings so the test
$(n, 1)$ string worldsheet feels {\sl time-dependent} electric field
as the radion rolls down. These are distinguished features never
seen in tachyon rolling of unstable D-branes or in other radion
rolling involving NS5-branes or D-branes. Third, in the limit
\bear g^2_{\rm st} N \gg g^2_{\rm st} n \gg 1, \label{limit} \eear
height of the radion potential, set by the binding energy density,
is estimated to be order \footnote{Throughout the paper, we shall
adopt the convention setting $2\pi$ times the string scale squared
$2\pi \ell^2_{\rm st} \equiv 2\pi \alpha'$ to unity, and measure all
dimensionful quantities in this unit.
We will also suppress length of the compactified direction $L_1$ by
setting it to unity. These dimensionful parameters can be recovered
by rescaling worldsheet variables. We shall denote the D$p$-brane
tension as $\tau_p$.} of
\bear \Delta {\cal V} = {1 \over g^2_{\rm st} (n+N)}. \label{deltav}
\eear
This can be made arbitrarily small compared to the string scale by
taking both the compactified dimension fixed in string unit and
$g^2_{\rm st} n$ sufficiently large. As such, all the energy
released would be carried off by the radiation of the type IIB
supergravity fields.

We have organized this paper as follows. In section 2, we set out
the Born-Infeld action of the $(n, 1)$ string in the background of
$N$ F-strings.  In section 3, we study detailed dynamics of the
radion rolling. In section 4, extend the analysis by including
worldsheet modulation of the bound-state string. In section 5, we
dwell on an exceptional situation, where the macroscopic F-strings
are probed by a supertube --- a polarized bound-state of D-particles
and F-strings.

\section{Setup}
\subsection{Supergravity Background of Macroscopic F-Strings}
Consider Type IIB string theory compactified on a circle
$\mathbb{S}_1$ of circumference $L_1$ of macroscopic size, $L_1 \gg
1$, and a closed string wrapped $N$ times around the circle
$\mathbb{S}_1$. Choosing the compactified coordinate to be $x^1$,
the $N$-times wrapped F-string gives rise to the (super)gravity
background in string frame as \cite{dabholkarharvey}
\bear \d s^2 &=& {1 \over H(r)} \Big( - \d t^2 + \d x_1^2 \Big) +
\d {\bf x}^2 \nonumber \\
B_{01} &=& {1 \over H(r)} - 1 \nonumber \\
e^{2 \phi} &=& g_{\rm st}^{2} {1 \over H(r)} \label{background}
\eear
where ${\bf x}$ denotes spatial coordinates transverse to the
macroscopic string, $r \equiv \vert {\bf x} \vert$, and the
harmonic function $H({\bf x})$ solving the transverse Laplace
equation is given in the supergravity regime by
\bear H(r) = \Big( 1 + {g^2_{\rm st} N \over r^6} \Big) \qquad \mbox{where}
\qquad  g^2_{\rm st} N \gg 1. \label{harmonic} \eear
Compared to other $p$-branes in string theory
\cite{horowitzstrominger}, the background exhibits several
distinguishing features. First, as one zooms into near horizon, ${r}
\rightarrow 0$, both the string coupling $e^{\phi(r)}$ and the NS-NS
$B$-field decrease monotonically: the string coupling interpolates
from $g_{\rm st}$ to $0$, and the electric component of the
$B$-field interpolates from $0$ to $-1$. This implies that, in the
supergravity background Eq.(\ref{background}), D-brane dynamics is
accessible in string perturbation theory and describable entirely in
terms of noncritical open strings \cite{ncos, klebanovmaldacena,
reyvonunge} near the horizon of the macroscopic F-strings.

\subsection{Dirac-Born-Infeld Action of $(p,q)$ String}
We are interested in the motion of $(p,q)$ string in the background
of the macroscopic F-string. For simplicity and definiteness, we
shall take $(p,q) = (n,1)$, viz. bound-state of $n$ F-strings and a
single D-string. We shall also take the F-string charge $n$ to be in
the range $1 \ll g^2_{\rm st} n \ll g^2_{\rm st} N$ for reasons that
will become clearer momentarily. At low-energy, dynamics of the
bound-state string is describable by Dirac-Born-Infeld (DBI) action
in the background of Eq.(\ref{background}). Denote the bound-state
string worldsheet coordinates as $\sigma^m$ $(m=0,1)$, and the
worldsheet fields as $X^M$ $(M=0,1, \cdots, 9)$ and ${\cal A}_m$
$(m=0,1)$. The DBI action then reads
\bear S_{\rm DBI}=- \tau_1 \int \d^2\sigma
\,e^{-(\phi-\phi_{\infty})}\sqrt{-\det(X^*(G+B)_{mn}+ {\cal
F}_{mn})}\quad, \label{dbiaction} \eear
where the pull-backs $X^*(G+B)$ are
\bear (X^*G)_{mn}&=&\frac{\partial
X^M}{\partial\sigma^{m}}\frac{\partial X^N}{\partial\sigma^{n}}
G_{MN}(X)\quad,\nonumber \\
(X^*B)_{mn}&=&\frac{\partial
X^M}{\partial\sigma^{m}}\frac{\partial
X^N}{\partial\sigma^{n}}B_{MN}(X)\quad. \eear
The DBI action is invariant under the reparametrization of the
worldsheet coordinates $\sigma^m \rightarrow \widetilde{\sigma}^m
(\sigma)$ and under the U(1) gauge transformation ${\cal A}_m
\rightarrow {\cal A}'_m = {\cal A}_m + \partial_m \epsilon$. We
shall fix the reparametrization invariance by choosing the static
gauge $\sigma^0=t=X^0$, $\sigma^1=X^1$, and fix the gauge
invariance by choosing the Coulomb gauge ${\cal A}_0 = 0$.

For the moment, we shall study homogeneous configurations in which
worldsheet fields are independent of $\sigma^1$. Because of the
Coulomb gauge chosen, ${\cal F}_{01}=\dot {\cal A}_1 \equiv {\cal
E}$ and the Gauss' law constraint
\bear \frac{\partial}{\partial\sigma^1}\left(\frac{\delta \CL_{\rm
DBI}} {\delta {\cal E}}\right)=0\quad \nonumber \eear
is trivially satisfied because of homogeneity along $\sigma^1$.
The DBI action is then reduced to
\bear S_{\rm DBI} \equiv \int \d t \d \sigma^1 {\cal L}_{\rm DBI}
= - \tau_1 \int \d t \d \sigma^1 \, \sqrt{\frac{1}{H}- \dot{\bf
X}^2- {1 \over H} \Big( (H- 1) -H {\cal E} \Big)^2}\quad.
\label{dbiexplicit} \eear

The momentum density ${\bf \Pi}_i$ conjugate to ${\bf X}^i$ is
given by
\bear {\bf \Pi}:= {\delta S_{\rm DBI} \over \delta \dot{\bf X}} =
\tau_1 \frac{\sqrt{H} \, \dot {\bf X}}{\sqrt{1 -H\dot {\bf
X}^2-\Big((H -1)- H {\cal E} \Big)^2}}\quad, \label{momentum}
\eear
The momentum density conjugate to ${\cal A}_1$ is given by the
displacement field:
\bear {\cal D} := {\delta S_{\rm DBI} \over \delta \dot A_1} = -
\tau_1 \frac{\sqrt{H} \Big( (H-1) - H {\cal E} \Big)} {\sqrt{1 -H
\dot {\bf X}^2- \Big( (H-1) - H {\cal E} \Big)^2}}\quad.
\label{displacement} \eear
The displacement field $\cal D$ is conserved and is directly
related to the F-string charge density inside the D-string. The
latter relation follows from the observation that in general
F-string current tensor $J^{MN}(x)$ couples minimally to the NS-NS
$B$-field:
\bear \Delta S_{\rm spacetime} = \int_{{\cal M}_{10}} \d^{10} x \,
B_{MN}(x) \Big( J^{MN}(x) + \cdots \Big) \label{minimal} \eear
where the ellipses abbreviate contributions from other spacetime
fields and, in particular, D-branes if present. Combining this with
the fact that the $B$-field couples to a D-brane through the
gauge-invariant combination $(X^*B + {\cal F})_{mn}$ and that the
proportionality constant in the minimal coupling Eq.(\ref{minimal})
is set by the F-string tension (which we set to unity), one finds
that the F-string charge density $J^{01}$ agrees precisely with the
displacement field Eq.(\ref{displacement}). The DBI action
Eq.(\ref{dbiaction}) describes dynamics of the $(p,q) = (n, 1)$
string, so the the F-string charge $n$ should be identified with the
displacement field $\cal D$:
\bear {\cal D} =n\quad. \nonumber \eear
%
%%%%%%%%%%%%%%%%%%%%%%%%%%%%%%%%%%%%%%%%%%%%%%%%%%%%%%%%%%%%%%%%%%%%%%%%%%%%%%%%%%%%%%%%%%%%
\subsection{Energy Density and Radion Potential}
From the canonical momenta, the Hamiltonian density ${\cal H}$ is
given by
\bear \CH&:=& {\bf \Pi} \cdot \dot {\bf X} + {\cal D} \, \dot{\cal
A}_1 - {\cal L}_{\rm DBI}
\nonumber \\
&=& \tau_1 \frac{1}{\sqrt{H}} \frac{1 - (H - 1) \Big( (H-1) - H
{\cal E} \Big) } {\sqrt{1- H \dot{\bf X}^2- \Big( (H -1) - H {\cal
E} \Big)^2}} \, . \label{H} \eear
For homogeneous configuration, the canonical momenta and the time
derivative of field variables are related by
\bear {\bf \Pi}^2 &=& \tau_1^2 \frac{H \, \dot{\bf X}^2 } {1 - H
\dot{\bf X}^2 - \Big((H - 1) - H {\cal E} \Big)^2}
\nonumber \\
{\cal D}^2 &=& \tau_1^2 \frac{ H \, \Big( ( H - 1) - H {\cal E}
\Big)^2} {1 - H \dot{\bf X}^2 - \Big((H - 1) - H {\cal E} \Big)^2}
\quad. \eear
These relations facilitate simplifying the Hamiltonian density
into the following forms:
 \bear {\cal H}
&=&{1 \over H}\left({\sqrt{ {H \tau_1^2} + {\cal D}^2} \over
\sqrt{1- H \dot{\bf X}^2}}
+  (H -1) {\cal D} \right)\nonumber\\
&=& {1 \over H} \left( \sqrt{H \Big(\tau_1^2 + {\bf \Pi}^2 \Big) +
{\cal D}^2}+  (H - 1) {\cal D} \right)\quad. \label{tension} \eear

Though the result Eq.(\ref{tension}) is elementary, one can
extract a lot of physics intuitions. We enlist some of them here.
\hfill\break
$\bullet$ Consider two extreme limits of F-string charge $n$. For
nearly D-string case, the energy density scales as $\tau_1$ times
$1/\sqrt{H}$. This is as it should be since the DBI action is
proportional to $e^{- \phi} \sqrt{G_{00} G_{11}} \sim 1/\sqrt{H}$.
For nearly F-string, the energy density scales as ${\cal D}$.
Again, this is as it should be since the first term's leading
contribution is $H^{-1} \, {\cal D}$ and cancels off part of the
second term's contribution $-(H^{-1}-1) \, {\cal D}$, thus scaling
as ${\cal D}$.

\noindent $\bullet$ Static energy of the $(n, 1)$-string is obtained from
Eq.(\ref{tension}) by setting ${\bf \Pi} = 0$, which we will
refer as `radion potential'.
As shown in Fig.1, the radion potential is a monotonic
function of $r$, equivalently,
$H(r)$.
%%%%%%%%%%%%%%%%%%% Fig. reduction %%%%%%%%%%%%%%%%%%%%%%%%%%%%
\begin{figure}
\epsfxsize=13cm \centerline{\epsfbox{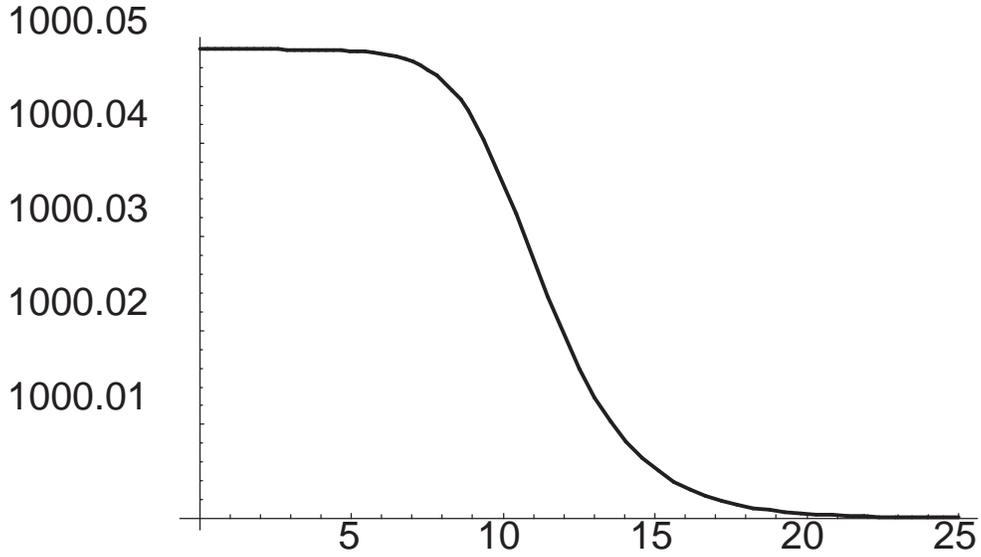}} \caption{\sl The
static radion potential of $(n, 1)$-string in the background of $N$
macroscopic F-string for $g_{\rm st} = 0.1$ and $n = 1000$. The
horizontal axis is $\log H(r)$ and the vertical axis is the static
energy density measured in unit of the F-string scale. The binding
energy of $(n, 1)$-string (as measured by the difference of the two
asymptotic plateau values) is parametrically small, and is about $5
\times 10^{-5}$ of the original/final energy. Radion rolls down the
potential hill and release the binding energy either into `radion
matter' or into radiation of supergravity modes.} \label{fig:pot}
\end{figure}
%%%%%%%%%%%%%%%%%%%%%%%%%%%%%%%%%%%%%%%%%%%%%%%%%%%%%%%%
At asymptotic infinity $r=\infty, H(r) = 1$, the energy density is
given by $\sqrt{\tau_1^2 + {\cal D}^2}$. As the center is
approached $r \rightarrow 0, H(r) \rightarrow \infty$, the energy
density scales to ${\cal D}$. The monotonic behavior demonstrates
that the $(n, 1)$ string can lower its energy density simply by
sliding down to the center, viz. toward the macroscopic F-strings.
Once the $(n, 1)$-string meets the macroscopic F-string, it will
form a non-threshold bound-state of $(n+N, 1)$-string. In the
limit $g^2_{\rm st} N \gg g^2_{\rm st}
n \gg 1$, the binding energy of a D-string to $(N+n)$
multiply wound F-string is extremely small, since
\bear \sqrt{\tau_1^2 + (N + n)^2} = \left[ (N+n) + {1 \over 2}
{1\over g_{\rm st}^2} {1 \over (N+n)} + \cdots \right] \, ,
\nonumber \eear
and the second term is of order ${\cal O}(1/g_{\rm st}^2 (N+n))$ and can be made
sufficiently small by taking the limit
\bear g_{\rm st}^2 (N + n) \gg 1 \qquad \mbox{and} \qquad g_{\rm
st} \ll 1 \quad. \label{binding condition} \eear
Note that this remains true for any length of the compactified
dimension as long as it is finite. Once the $(n,1)$ string binds
with the background $N$ F-strings, the released binding energy
$\Delta E$ would be
\bear \Delta E &=& \Big\{ \Big[N + \Big( n + {1 \over 2} {1 \over
g_{\rm st}^2} {1 \over n} + \cdots \Big) \Big] - \Big[(N + n) + {1
\over 2}
{1 \over g_{\rm st}^2} {1 \over N + n} + \cdots \Big]\Big\} \nonumber \\
&\sim& {\cal O} \Big( {1 \over g_{\rm st}^2 n} \Big) \quad, \eear
thus setting the height of the radion static potential,
Eq.(\ref{deltav}). We can render the total released energy
sufficiently small compared to the string scale and the dynamics
perturbative by taking the limit
\bear g_{\rm st}^2 n \gg 1 \qquad \mbox{and} \qquad g_{\rm st} \ll
1 \qquad. \label{release condition} \eear
Therefore, under this circumstance, emission of massive closed
string states would be kinematically forbidden, and the entire
process of the radion rolling of $(n,1)$-string in the macroscopic
F-string background is amenable completely within the regime of
low-energy supergravity approximation!

Combining the two conditions Eqs.(\ref{binding condition},
\ref{release condition}), we are led to consider the limit alluded
already in the Introduction.

\noindent $\bullet$ One novel aspect of the radion rolling process
is that the worldvolume electric field is not a constant but ought
to change simply because the pullback of NS-NS $B-$field varies
radially and because the displacement is already fixed by the
F-string charge. To see how the electric field behaves, consider
the displacement field near the asymptotic infinity and the
center. Near the asymptotic infinity, $r \rightarrow \infty$ and
$H \rightarrow 1$, the displacement field scales as
\bear {\cal D} \, \longrightarrow \, {1 \over g_{\rm st}} {{\cal
E}_\infty \over \sqrt{1 - {\cal E}_\infty^2}}.
\label{displacement2} \eear
Thus, the electric field on the D-string sitting at asymptotic
infinity would take a value near the critical field ${\cal E}_\infty
\sim {\cal O} (1)$. This value (hence the critical electric field
minus this value) is set by the net F-string charge. Now, as the
radion rolls down toward the macroscopic F-string, the electric
field gets increased and attain precisely the critical value at the
moment the $(n, 1)$-string hits the macroscopic F-string. This is
readily seen from Eqs. (\ref{displacement}, \ref{H}): the
displacement field may be expressed as
\bear {D \over H^2} =- \left({\cal H}-{\cal D}(1-H^{-1})\right)
\left((1-H^{-1})-{\cal E}\right),
\nonumber \eear
so, in case the displacement field ${\cal D}$ is uniform on the
worldsheet, the electric field approaches the critical value $1$ as
\bear {\cal E} \rightarrow \Big( 1  - {\cal O} ({1 \over H})
\Big). \eear
It shows that the $(n, 1)$-string turns into a {\sl tensionless}
noncritical string as the radion rolls down to the macroscopic
F-strings \footnote{Actually, the conclusion is more general and
holds also for $n=0$.}.

%%%%%%%%%%%%%%%%%%%%%%%%%%%%%%%%%%%%%%%%%%%%%%%%%%%%%%%%%%%%%%%%%%%%%
\subsection{Constraints and Conserved Quantities} Before
proceeding further, we shall discuss conserved quantities during the
homogeneous radion rolling process. By construction, the DBI action
is invariant under the worldsheet reparametrization and gauge
transformation. In the previous section, we have fixed them by
choosing the static gauge for the reparametrization invariance and
the Coulomb gauge for the gauge invariance. As such, the dynamics
described in terms of the gauge-fixed DBI action ought to be
supplemented by the corresponding constraints. Following the methods
originally developed in \cite{reyyee}, we now demonstrate that these
constraints lead to conserved charges.

For the gauge invariance, the constraint gives rise to the Gauss'
law:
\bear \partial_\sigma \Big( {\delta S_{\rm DBI} \over \delta {\cal
E}} \Big) =
\partial_\sigma {\cal D} = 0. \nonumber \eear
It implies that the displacement field is constant-valued along
the $(1,n)$-string, and is a conserved quantity. For homogeneous
worldsheet field configurations, this is automatically satisfied.

For the reparametrization invariance, the constraints are
obtainable by taking the DBI action {\sl prior} to the
gauge-fixing
\bear S_{\rm DBI} = - \tau_1 \int \d t \d \sigma \, \sqrt{-H
({\cal G}_{00} {\cal G}_{11} - {\cal G}_{01}{\cal G}_{10}) },
\label{dbifullaction} \eear
where
\bear {\cal G}_{00} &:=& {1 \over H} \Big( -{\dot X}^2_0 + {\dot
X}_1^2 \Big) + \dot{\bf X}^2 \nonumber \\
{\cal G}_{11} &:=& {1 \over H} \Big(- X_0^{'2} + X_1^{'2} \Big) +
{\bf X}^{'2} \nonumber \\
{\cal G}_{01} &:=& -({1 \over H} - 1) \Big(\dot{X}_0 X_1' - X_0'
\dot{X}_1 \Big) + \frac{1}{H}\left(-\dot{X}_0 X_0'+\dot{X}_1
X_1'\right)+\dot {\bf X} \cdot {\bf X}' + {\cal
E}\nonumber\\
{\cal G}_{10} &:=& + ({1 \over H} - 1) \Big(\dot{X}_0 X_1' - X_0'
\dot{X}_1 \Big) + \frac{1}{H}\left(-\dot{X}_0 X_0'+\dot{X}_1
X_1'\right)+ \dot {\bf X} \cdot {\bf X}' -{\cal E} \eear
and deriving the field equations of motion for $X^0=-X_0$ and
$X^1$.

For $X^0$, the equation of motion reads
\bear
\partial_t\, T^{00}+\partial_\sigma T^{10}=0\quad,
\nonumber \eear
where
\bear
T^{00}&=&\frac{\tau_1\left[-\dot{X}_0\CG_{11}+\frac{1}{2}(H-1)X_1'(\CG_{01}-\CG_{10})+\frac{1}{2}
X_0'(\CG_{01}+\CG_{10})\right]}{\sqrt{-H({\cal G}_{00} {\cal G}_{11}
- {\cal G}_{01}{\cal
G}_{10})}}\nonumber\\
T^{10}&=&\frac{\tau_1\left[-X_0'\CG_{00}-\frac{1}{2}(H-1)\dot{X}_1(\CG_{01}-\CG_{10})+\frac{1}{2}
\dot{X}_0(\CG_{01}+\CG_{10})\right]}{\sqrt{-H({\cal G}_{00} {\cal
G}_{11} - {\cal G}_{01}{\cal G}_{10})}}\quad. \nonumber \eear

In the static gauge $t=X^0$, $\sigma=X^1$, they become
\bear {\cal H} &=& \frac{\tau_1}{\sqrt{H}} {\left[1 + H {\bf X}^{'2}
- (H - 1) \left( (H-1) - H {\cal E} \right) \right] \over \sqrt{(1 -
H \dot{\bf
X}^2)(1+H{\bf X}^{'2}) - \left( (H-1) - H{\cal E} \right)^2+H^2(\dot{\bf X}\cdot {\bf X}^{'})^2}}\nonumber\\
T^{10}&=&{\tau_1 \over \sqrt{H}} { -H \, (\dot{\bf X}\cdot {\bf
X}^{'})\over \sqrt{(1 - H \dot{\bf X}^2)(1+H{\bf X}^{'2}) - \left(
(H-1) - H{\cal E} \right)^2+H^2(\dot{\bf X}\cdot {\bf
X}^{'})^2}}\quad. \nonumber \eear
They are nothing but the canonical energy density and the momentum
density.

Repeating the analysis for $X^1$, the equation of motion is
\bear
\partial_t\, T^{01}+\partial_\sigma T^{11}=0\quad, \nonumber
\eear
where
\bear
T^{01}&=&\frac{\tau_1\left[\dot{X}_1\CG_{11}-\frac{1}{2}(H-1)X_0'(\CG_{01}-\CG_{10})-\frac{1}{2}
X_1'(\CG_{01}+\CG_{10})\right]}{\sqrt{-H({\cal G}_{00} {\cal G}_{11}
- {\cal G}_{01}{\cal
G}_{10})}}\nonumber\\
T^{11}&=&\frac{\tau_1\left[X_1'\CG_{00}+\frac{1}{2}(H-1)\dot{X}_0(\CG_{01}-\CG_{10})-\frac{1}{2}
\dot{X}_1(\CG_{01}+\CG_{10})\right]}{\sqrt{-H({\cal G}_{00} {\cal
G}_{11} - {\cal G}_{01}{\cal G}_{10})}}\quad. \nonumber \eear
Again, in the static gauge, they are
\bear T^{01}&=&- {\tau_1 \over \sqrt{H}} { H \, (\dot{\bf X}\cdot
{\bf X}^{'})\over \sqrt{(1 - H \dot{\bf X}^2)(1+H{\bf X}^{'2}) -
\left( (H-1) - H{\cal E} \right)^2+H^2(\dot{\bf X}\cdot {\bf
X}^{'})^2}}
\nonumber\\
{\cal P} &=&-\frac{\tau_1}{\sqrt{H}} {\left[1 - H \dot{\bf X}^{2} -
(H - 1) \left( (H-1) - H {\cal E} \right) \right] \over \sqrt{(1 - H
\dot{\bf X}^2)(1+H{\bf X}^{'2}) - \left( (H-1) - H{\cal E}
\right)^2+H^2(\dot{\bf X}\cdot {\bf X}^{'})^2}}\quad, \label{press1}
\eear
which are the momentum density and the pressure.

%%%%%%%%%%%%%%%%%%%%%%%%%%%%%%%%%%%%%%%%%%%%%%%%%%%%%%%%%%%%%%%%%%%%%%%%%%%%%%%%%%%%%%%%%%%%%
\subsection{Pressure}
By the canonical method, we also obtain the pressure density on the
worldsheet of $(n, 1)$-string as
\bear {\cal P} &=&{\partial L_{\rm DBI} \over \partial {\bf X}'}
\cdot {\bf X}' + {\cal D} (- \dot {\cal A}_1)+ {\cal L}_{\rm DBI}
\nonumber \\
&=&-{1 \over H} \left[ \sqrt{ \Big(H \tau_1^2 + {\cal D}^2 \Big)
\Big( 1 - H \dot{\bf X}^2 \Big)} +  (H - 1) {\cal
D} \right] \nonumber \\
&=& - {1 \over H} \left[ {\left(H \tau_1^2 + {\cal D}^2\right) \over
\sqrt{H \Big(\tau_1^2 + {\bf \Pi}^2 \Big) + {\cal D}^2}} + (H - 1)
{\cal D} \right] \quad, \label{pressure} \eear
Notice that the pressure density always remains negative-definite.

In case of unstable D-brane, a salient feature was that the endpoint
of the tachyon rolling comprises of pressureless `tachyon matter'.
Given that the radion rolling process can be mapped to an analog
tachyon rolling, the pressure Eq.(\ref{pressure}) may analogously
reduce to a pressureless endpoint constituents. This turns out
roughly the case, as we will analyze in detail momentarily, except
that the endpoint constituents still carries the pressure exerted by
the $n$ F-string constituents albeit reduced from the initial value.

To analyze the rolling behavior of the pressure, we begin with the
conserved energy density. From Eq.(\ref{tension}), one readily
finds that
\bear \sqrt{H \Big(\tau_1^2 + {\bf \Pi}^2\Big) + {\cal D}^2} = H
\, {\cal H} - (H - 1) {\cal D}. \nonumber \eear
Utilizing the relation in Eq.(\ref{pressure}), the pressure can be
recast in terms of conserved quantities as
\bear {\cal P} = - \Big(1 - {1 \over H} \Big) {\cal D} - {1 \over
H} \Big({H \tau_1^2 + {\cal D}^2 \over H ({\cal H} - {\cal D}) +
{\cal D}} \Big). \label{Pexp} \eear
It shows the following characteristic crossover scale for $H$
\bear H_\star \equiv \Big({{\cal D} \over \tau_1} \Big)^2 \simeq
g_{\rm st}^2 n^2 \gg 1. \nonumber \eear
Across the scale $H(r) \sim H_\star$, the numerator of the last
term turns over the dominant contribution from either D-string or
F-string constituents. The same argument holds for the denominator
of the last term once we make use of the limit $g_{\rm st}^2 n \gg
1$ in Eq.(\ref{Pexp}) and approximate ${\cal H} - {\cal D}$ as $1/
%\over
(2%} (1 /
g_{\rm st}^2 {\cal D})%)
+ \cdots$. Built upon these
observations, one can re-express the pressure as
\bear {\cal P} = - {\cal D} - \Big( {H_\star \over H + 2 H_\star}
\Big) {1 \over g_{\rm st}^2 {\cal D}} + \cdots. \nonumber \eear
As the radion rolls down, viz. the $(n,1)$-string is attracted to
the macroscopic F-string, the function $H(r)$ varies monotonically
from 1 to $+\infty$. Accordingly, the coefficient of the second
term (inside the parenthesis) decreases monotonically from $1/2$
to $0$, demonstrating transparently that the pressure of the
$(n,1)$ string decreases from $- \sqrt{\tau_1^2 + {\cal D}^2}$ at
infinite separation from the macroscopic F-string to $- {\cal D}$
at the horizon.

The observation that the pressure exhibits a crossover around $H(r)
\sim H_\star$ and form a radion matter of reduced pressure can be
seen more intuitively as follows. Consider the difference between
the energy density and the pressure density and utilize the Schwarz
inequality:
\bear {\cal H} - {\cal P} &=& {1 \over H} \left[ 2(H - 1) {\cal D}
+ \sqrt{ H (\tau_1^2 + {\bf \Pi}^2) + {\cal D}^2} + {( H \tau_1^2
+ {\cal D}^2) \over \sqrt{H (\tau_1^2 + {\bf \Pi}^2) + {\cal
D}^2}}
\right] \nonumber \\
&\ge& 2 \left[ {\cal D} + {1 \over H} \Big( -{\cal D} + {\cal D}
\sqrt{{H \over H_\star} + 1} \Big) \right]. \eear
In the asymptotic region $r\rightarrow \infty$, equivalently,
$H(r) \rightarrow 1$, the quantity in the last line yields $2
\sqrt{g_{\rm st}^{-2} + {\cal D}^2}$. This is nothing but twice of
the $(n, 1)$-string tension. Recall that $(n, 1)$-string is a BPS
configuration, thus the energy density and the pressure takes the
same value. Therefore the Schwarz inequality is actually saturated
in the asymptotic region. Rolling inward, one readily observe that
the inequality remains saturated in the region where $1 \ll H(r)
\ll H_\star$. The quantity in the last line now yields
\bear {\cal H} - {\cal P} \quad &\rightarrow& 2 \Big[ \Big( {\cal
D} - {{\cal D} \over H} + \cdots \Big) + \Big( {{\cal D} \over H}
+ {\tau_1^2 \over 2} {1 \over {\cal D}} + \cdots
\Big)\Big] \nonumber \\
&=& 2 \Big[ {\cal D} + {1 \over 2 g_{\rm st}^2 } {1 \over {\cal
D}} + \cdots \Big], \label{first} \eear
and is readily recognized as twice of the energy of $(1,
n)$-string. Taking into consideration of the constraint $|{\cal
P}| \le {\cal E}$ due to Lorentz invariance on the string
worldsheet and of the conserved energy, this implies that the
pressure has not yet relaxed in the region $1 \ll H(r) \ll
H_\star$. Rolling further into the near-coincidence region $H(r)
\gg H_\star$, the quantity in the last line approaches $2 {\cal
D}$ from above:
\bear {\cal H} - {\cal P} \quad &\rightarrow& \quad 2 \Big[ {\cal
D} + {1 \over g_{\rm st} \sqrt{H}} + \cdots \Big]. \label{second}
\eear
As the energy remains conserved throughout, by comparing the two
expressions Eq.(\ref{first}) and Eq.(\ref{second}), one learns
that the pressure is lowered as the $(n,1)$-string approaches to
the macroscopic F-strings.

%%%%%%%%%%%%%%%%%%%%%%%%%%%%%%%%%%%%%%%%%%%%%%%%%%%%%%%%%%%%%%%%%%%
\section{Radion Dynamics for Nonzero Angular Momentum}
%%%%%%%%%%%%%%%%%%%%%%%%%%%%%%%%%%%%%%%%%%%%%%%%%%%%%%%%%%%%%%%%%%%
\subsection{Effective Potential}
In general, the rolling radion dynamics between the $(n,1)$-string
and the macroscopic $N$ F-string would involve nonzero impact
parameter and hence angular momentum. In this section, we shall
study such situations in detail. Because of SO(8) symmetry in the
transverse space of ${\bf X}$'s, the trajectories of the two-body
motion under the SO(8) invariant central potential always lie on a
plane. We shall take the plane to be along
$(X^8=R\cos\Theta,X^9=R\sin\Theta)$. Again, we focus on
homogeneous worldsheet field configurations. By repeating the
canonical analysis as in section 2, one readily obtains conserved
quantities such as energy ${\cal H}$ and angular momentum $\ell$.
By SO(8) symmetry, the angular momentum $\ell$
\bear \ell = (R^2 \dot \Theta) \frac{\sqrt{H \tau_1^2 + {\cal
D}^2}}{\sqrt{1 - H(\dot R^2 + R^2 \dot \Theta^2)}} \nonumber \eear
is a conserved quantity. So, for the motion with nonzero angular
momentum $\ell$, the energy density ${\cal H}_\ell$ for motion
with a fixed angular momentum $\ell$ is given by
\bear \CH_\ell &=& {1 \over H} \left[ \frac{\sqrt{ H \tau_1^2 +
{\cal D}^2}}{\sqrt{1 - H (\dot{R}^2 + R^2 \dot{\Theta}^2})} + (H -
1) {\cal D} \right]
\nonumber \\
&=& {1 \over H} \left[ \sqrt{ H \Big(\tau_1^2 + {\Pi_r}^2 +
{\ell^2 \over R^2}\Big) + {\cal D}^2} + (H - 1) {\cal D}\right],
\label{Eangular}\eear
where $\Pi_r$ is the conjugate momentum to $R$. It shows that
nonzero angular momentum gives rise to the contribution of the
centrifugal barrier through the first term. So, for motions with a
given angular momentum $\ell$, the effective potential is obtained
by setting ${\Pi_r}=0$ in ${\cal H}$. Schematic view of the
effective potential is plotted in Fig.2.
%%%%%%%%%%%%%%%%%%% Static energy with angular momentum  %%%%%%%%
\begin{figure}
\epsfxsize=13cm \centerline{\epsfbox{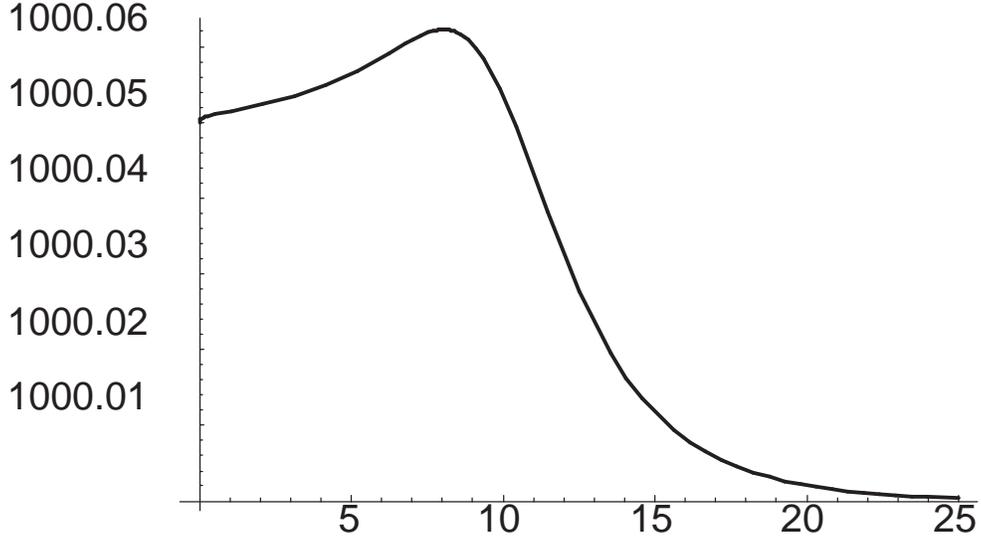}}
\caption{\sl The
effective potential of $(n, 1)$-string circulating the background of
$N$ macroscopic F-string for $g_{\rm st} = 0.1$, $n = 1000$ and
$\ell^2/(g_{\rm st}^2 N^2) = 1.25$. The horizontal axis is $\log
H(r)$ and the vertical axis is the static energy density measured in
unit of F-string tension $\tau_{\rm F}$. Compared to the static
potential at zero angular momentum, there shows up an angular
momentum-dependent potential barrier. As the angular momentum $\ell$
is increased, the barrier gets higher.} \label{fig:potl2}
\end{figure}
%%%%%%%%%%%%%%%%%%%%%%%%%%%%%%%%%%%%%%%%%%%%%%%%%%%%%%%%
It shows that the angular momentum gives rise to a potential barrier between the
asymptotic region and the near-center region. Since the initial radial velocity increases
the energy, the effective potential implies that up to a critical radial velocity, the
$(n,1)$-string would experience centrifugal barrier. Beyond that critical velocity, the
motion would simply fall into the near-center region.

\subsection{Pressure and Angular Momentum}
In the previous section, we have seen that the pressure on the
$(n,1)$-string worldsheet relaxes steadily as the string moves
toward the macroscopic F-string. A pertinent question is to explore
possible effects of nonzero angular momentum to the pressure. A
priori, such effects might be of quite a complicated nature much the
way the monotonic static potential in Fig.1 was turned by nonzero
angular momentum into an effective potential with features shown in
Fig. 2.

In this subsection, we will show that, once the string is captured
to the macroscopic F-string, the $(n,1)$-string diminishes the
pressure in a universal manner irrespective of the angular momentum.
In other words, the relaxation process of the $(n,1)$-string with
nonzero angular momentum is the same as that of the $(n,1)$-string
with vanishing angular momentum. This is quite an interesting result
and implies that nonzero angular momentum does not exert a barrier
to the pressure, in sharp contrast to the energy.

By the canonical method, the pressure ${\cal P} = T_{11}$ is
obtained as before. The result for nonzero angular velocity is
\bear {\cal P} = -{1 \over H} \Big[\sqrt{ \Big(H \tau_1^2 + {\cal
D}^2\Big) \Big( 1 - H (\dot{R}^2 + R^2 \dot\Theta^2)\Big)} + (H-1)
{\cal D} \Big]. \nonumber \eear
The result is as expected --- it simply expresses the radial and
the azimuthal motion in two-dimensional plane on which the motion
lies. Using the relations
\bear \Pi^2_r &=& \tau_1^2 {H \dot{R}^2 \over 1 - H
(\dot{R}^2 + R^2 \dot{\theta}^2 ) - \Big((H-1) - H {\cal E} \Big)^2} \nonumber \\
{\ell^2 \over R^2} &=& \tau_1^2 {H \, R^2 \dot{\theta}^2 \over 1 - H
(\dot{R}^2 + R^2 \dot{\theta}^2 ) - \Big((H-1) - H {\cal E}\Big)^2},
\nonumber \eear
one readily finds that the pressure for the motion with angular
momentum $\ell$ is given in terms of conserved quantities as
\bear {\cal P}_\ell &=& -{1 \over H} \Big[ (H-1) {\cal D} + {H
\tau_1^2 + {\cal D}^2 \over \sqrt{H \Big(\tau_1^2 + \Pi_r^2 +
{\ell^2 \over R^2} \Big) + {\cal D}^2}} \Big]. \label{Pangular}
\eear
Comparing the expression with the conserved energy
Eq.(\ref{Eangular}), one finds that influence of the angular
momentum on the pressure is the same as that on the energy ---
inside the square-root, ${\bf \Pi}^2$ is replaced by $(\Pi_r^2 +
\ell^2/R^2)$. Therefore, as for the situation of zero angular
momentum, by eliminating the square-root in favor of the conserved
energy ${\cal H}_\ell$, we obtain
\bear {\cal P}_\ell = - \Big( 1 - {1 \over H}\Big) {\cal D} -
{\Big(\tau_1^2 + {1 \over H}{\cal D}^2\Big) \over H ({\cal H}_\ell
- {\cal D}) + {\cal D}}. \nonumber \eear
Since the above functional form is precisely the same as that for
zero angular momentum once ${\cal H}, {\cal P}$ are replaced by
${\cal H}_\ell, {\cal P}_\ell$, the relaxation process of the
pressure would be universal, independent of the value of the
angular momentum.

\subsection{Capture Cross-Section}
Now that planar radion dynamics is governed by the effective
potential exhibiting potential barrier, we shall estimate the
cross-section for the $(n,1)$-string being captured by the
macroscopic F-string. Let the $(n,1)$-string impinge on the
macroscopic F-string at $r\rightarrow\infty$ with the impact
parameter $L$ and the radial velocity $V$. From the expression for
$\CE$ and $\ell$, it is straightforward to see, as
$r\rightarrow\infty$, that the conserved quantities are
\bear \CH\rightarrow\frac{\sqrt{\tau_1^2 + {\cal
D}^2}}{\sqrt{1-V^2}}\qquad \mbox{and} \qquad
\ell\rightarrow\frac{\sqrt{\tau_1^2 + {\cal D}^2}}{\sqrt{1-V^2}}\,
LV\quad. \eear
Notice that, for such motions, the energy is always larger than
the rest mass $\sqrt{\tau_1^2 + {\cal D}^2}$.

How does the effective potential, as depicted in Fig.2, scale with
the angular momentum $\ell$? As the $(n,1)$-string moves inward,
the effect of nonzero angular momentum scales inversely with the
radial distance-squared, and hence dominates over the effect of
the harmonic function $H(r)$. Denote the location of the maximum
of the effective potential as $r_\star$ and $H_\star \equiv
H(r_\star)$. The effective potential behaves in ways that, as the
angular momentum $\ell$ is cranked up, the maximum location
$r_\star$ and $H_\star$ is pushed outward, and height of the
potential barrier $V_{\rm eff} (H_\star) - V_{\rm eff} (0)$ is
increased.

Such behavior of the effective potential entails qualitative
features that, for a given angular momentum $\ell$, there exists a
critical energy $\CH_{\rm cr} (\ell)$ such that the motion
starting at infinity with $\CH<\CH_{\rm cr}$ would be bounced at
some finite radial distance back to infinity. Functional form for
$\CH_{\rm cr} (\ell)$ as a function of $\ell$ (as well as other
quantum numbers $n, N$) is rather complicated. Fortunately, its
explicit functional form would not be needed, as it is evident
that $\CH_{\rm cr} (\ell)$ increases monotonically with $\ell$
simply because the effective potential itself increases
monotonically by the contribution of nonzero angular momentum
$\ell$. Being monotonic, the relation implies that $\CH_{\rm cr}
(\ell)$ is invertible to a function $\ell_{\rm cr} \equiv \ell
(\CH_{\rm cr})$. Turning around the arguments, since we have a
well-defined inverse $\ell_{\rm cr} (\CH)$, the motion would be
bounced off if $\ell>\ell_{\rm cr}(\CH)
$ for a given $\CH$. Given an
initial velocity $V$, it happens when the impact parameter $L$ is
large enough:
\bear L\,>\,\frac{V^{-1} \sqrt{1-V^2}}{\sqrt{\tau_1^2 + {\cal
D}^2}}\,\, \ell_{\rm cr}(\CH_\ell) \qquad \mbox{where} \qquad
\CH_\ell =\frac{\sqrt{\tau_{ 1}^2+ {\cal
D}^2}}{\sqrt{1-V^2}}\quad. \nonumber \eear

For motions with the impact parameter less than $\ell_{\rm cr}(\cal H)$,
the motion of the $(n,1)$-string would reach all the way to the
macroscopic F-string located at $r=0$. In the previous section, we
have shown that the pressure relaxes steadily as the radion rolls
to the center $r=0$, and eventually the $(n,1)$-string forms a
bound-state with the macroscopic F-string. Therefore, we can
define a total cross-section $\sigma_{\rm capture}$ for the
$(n,1)$-string being captured by the macroscopic F-string as
\bear \sigma_{\rm capture}(V)\,=\, \frac{16\pi^3}{105}
\left(\frac{V^{-1} \sqrt{1-V^2}}{\sqrt{\tau_{1}^2+ {\cal D}^2}}\,\,
\ell_{\rm cr}(\CH_\ell) \right)^7, \nonumber \eear
where $\frac{16\pi^3}{105}$ is the angular volume of the
transverse disc $\mathbb{S}^7$. For the scattering states, in
principle it is possible to get the differential cross-section,
but because analytic solution for the motions is not in our hand,
it is not very illuminating to analyze further.

\section{Mapping to Rolling of `Analog' Tachyon}
We have seen in the above analysis that a $(n,1)$ string moving in
the background of large-$N$ F1 strings generically falls toward
$r\rightarrow 0$ if angular momentum is not too large. As emphasized
by Kutasov \cite{kutasov} and further studied in subsequent works
\cite{followup}, the situation bears a certain similarity with
rolling tachyon dynamics of unstable D-branes. In this section, we
shall push such analogy to our rolling radion dynamics in a more
quantitative manner.

Begin with the radial dynamics as described by Eq.(\ref{Eangular}),
\be \dot R^2\,=\,{1 \over H}\left[1-\frac{\left(\tau_1^2 +
\frac{\ell^2}{R^2}\right)H+ {\cal D}^2} {\left( H (\CH - {\cal D})
+ {\cal D}\right)^2}\right]\quad. \label{rdot} \ee
As we discussed in the previous section, for $\CH$ and $\ell$
below threshold, the orbit of the bound-state string is bound to
the horizon $R=0$. Near $R\sim 0$,
\bear \dot R^2\,\sim\,{1 \over H}\,\sim\,\frac{R^6}{g^2_{\rm st} N}\qquad
\mbox{so} \qquad
R\,\sim\,\frac{(g^2_{\rm st} N)^{\frac{1}{4}}}{\sqrt{2}}\,\,t^{-\frac{1}{2}}
\quad. \nonumber \eear
It shows that, to reach the horizon at $R=0$, the motion takes an
infinite coordinate time lapse, but a finite proper time lapse.

As discussed in the previous sections, dynamics with zero angular
momentum is described by one-dimensional motion, while those with
nonzero angular momentum is reduced to two-dimensional motion. When
mapped to analog tachyon rolling dynamics, these two situations
correspond to real-valued tachyon relevant for unstable D-brane
decay and to complex-valued tachyon relevant for ${\rm D}
\overline{\rm D}$ brane-antibrane pair. We now demonstrate such
correspondences.

\subsection{D-string with zero angular momentum}
Consider first the motion of a pure D-string (so that ${\cal D}=0$)
with no angular momentum $\ell = 0$. In such a situation, we can
throw away the complicated term in the action involving gauge field
and take a simple effective action,
\bear S_{\rm D1} = -\tau_{1} \int \d
t\,\frac{1}{\sqrt{H}}\,\sqrt{1- H \, \dot R^2}\quad. \nonumber
\eear
The motion is one-dimensional, so $R$ can be taken any one of the 8
scalar fields ${\bf X}$, say, $R = X^9$. We then introduce analog
'tachyon' field $T$, a real-valued scalar field, as the proper
variation in the target space:
\bear \sqrt{H} \d R \,\, \equiv \, -\d T\quad \rightarrow \quad T(R)
= -\int^R \d R' \sqrt{H(R')}. \label{tachyon}\eear
It shows that radion dynamics for zero angular momentum is analogous
to tachyon dynamics of unstable D-brane.

In the near-horizon region, $R \rightarrow 0$, the analog tachyon
field moves out to infinity:
\bear T(R)\,\,\sim\,\,\frac{\sqrt{g^2_{\rm st}
N}}{2}\frac{1}{R^2}\quad. \nonumber \eear
In terms of the tachyon field Eq.(\ref{tachyon}), the D-string
action is expressible as
\bear S_{\rm D1}\,\,=\,\,-\tau_{1} \int \d
t\,V(T)\,\sqrt{1-\dot{T}^2}\quad, \nonumber \eear
where the 'tachyon' potential is denoted as $V(T)$. Near the
horizon, $T\rightarrow \infty$, and the tachyon potential decays in
power-law type:
\bear V(T)\,\,\sim\,\,(g^2_{\rm st} N)^{1 \over 4} {1 \over (8
T^3)^{1 \over 2}} \quad. \label{zeroangular} \eear

Likewise, the pressure, $\cal P$, is expressible in terms of the
analog tachyon field $T$. Again, utilizing Eq.(\ref{rdot}) to
Eq.(\ref{pressure}), one finds that the pressure asymptotes as
\bear {\cal P}\, \sim \,-\frac{\tau_{1}^2}{g^2_{\rm st} N\CH} \,
R^6 \,\sim\,-\frac{\tau_{1}^2}{8 \CH} \sqrt{g^2_{\rm st} N} {1
\over T^3}\quad. \nonumber \eear
It shows that the pressure on the D-string worldsheet vanishes in
power-law. We therefore see that both the tachyon potential and the
pressure decay at $T \rightarrow \infty$ much like the tachyon
rolling of unstable D-branes except that the asymptotic behavior is
power-like rather than behaving exponentially.

\subsection{$(n,1)$-string with nonzero angular momentum}
We shall now extend the mapping to more general situations with
${\cal D}\neq 0$ and $\ell\neq 0$. Begin with the Hamiltonian
Eq.(\ref{tension}):
\bear \CH\,=\,\frac{1}{H}\left(\sqrt{H\left(\tau_{1}^2+{\bf
\Pi}^2\right) + {\cal D}^2}+ (H-1) {\cal D}\right)\quad. \nonumber
\eear
For the motion with nonzero angular momentum, we simply substitute
in the $(R,\Theta)$-plane as
\bear {\bf
\Pi}^2\,\,\rightarrow\,\,\Pi_r^2+\frac{\Pi_{\theta}^2}{R^2}\,\,=
\,\,\Pi_r^2+\frac{\ell^2}{R^2}\quad, \nonumber \eear
where $\Pi_r$ and $\Pi_{\theta}$ are conjugate momenta to $R$ and
$\Theta$, respectively. We shall obtain the effective Lagrangian
for the motion with a fixed angular momentum $\ell$ by performing
the Legendre transform with respect to $\Pi_r$. First solve
$\Pi_r$ in terms of $\dot R$ from
\bear \dot R\,\,=\,\,\frac{\partial \CH}{\partial
\Pi_r}\,\,=\,\,\frac{\Pi_r}{\sqrt{H\,
\left(\tau_{1}^2+\Pi_r^2+{\ell^2 \over R^2}\right)+ {\cal
D}^2}}\quad. \nonumber \eear
We then define effective Lagrangian $L_{\rm eff}$ by Legendre
transform:
\bear {\cal L}_{\rm eff}&=& \Pi_r \, \dot R -\CH\nonumber \\
&=& -\frac{1}{H}\left(\frac{H \, \Big(\tau_{1}^2+{ \ell^2 \over
R^2}\Big) + {\cal D}^2} {\sqrt{H \, \Big(\tau_{1}^2+\Pi_r^2+
{\ell^2 \over R^2}\Big) + {\cal D}^2}}+(H-1) {\cal D}\right)\nonumber\\
&=&-\frac{1}{H}\left(\sqrt{H\, \Big(\tau_{1}^2+{\ell^2 \over R^2}
\Big) +{\cal D}^2}\,\,%\cdot
\sqrt{\Big(1-H \, \dot R^2\Big)} +(H-1)
{\cal D}\right)\quad.\nonumber \eear

As discussed in previous sections, dynamics for nonzero angular
momentum as described by the above effective Lagrangian is
describable on a two-dimensional plane in ${\bf X}$-space. Take
the plane along $(8,9)$-directions and denote $Z \equiv (X^8+i
X^9)$. We then see that the proper variation of radial field $R
\equiv \vert Z \vert$ on the plane can be mapped to a
complex-valued tachyon field $T$ by:
\bear \sqrt{H} \,\d Z\,\,=\,\, -\d T \, , \qquad \sqrt{H} \, \d
\overline{Z} \, \, = \, \, - \d \overline{T} \, . \nonumber \eear
Utilizing the planar symmetry, the effective Lagrangian then takes
the form
\bear S_{\rm eff}\,=\,- \int \d t\,\left[\tau_{1} V_{\rm eff}
(T\overline{T})\,\sqrt{1-\vert{\dot T}\vert^2} +W_{\rm eff}
(T\overline{T}) \, {\cal D}\right]\quad, \nonumber \eear
where the effective tachyon potential $V_{\rm eff} (\vert T\vert)$
and the background potential $W_{\rm eff} (\vert T\vert)$ asymptote
at $\vert T \vert \rightarrow\infty$ to:
\bear V_{\rm eff} (\vert T\vert )\,\,\sim\,\,\frac{\ell}{\tau_{1}
\sqrt{g^2_{\rm st} N}}\,R^2\,\,\sim\,\, \frac{\ell}{2\tau_{1}} {1
\over \vert T\vert} \qquad \mbox{and} \qquad W_{\rm eff} (\vert
T\vert) \sim 1 \quad. \nonumber \eear
Compared to the dynamics with zero angular momentum
Eq.(\ref{zeroangular}), we learn that nonzero angular momentum
makes the tachyon potential vanishes slower.

The pressure can be worked out similarly. We found that
\bear {\cal P}\,\,\sim\,\,- {\cal D}+\left({\cal
D}-\frac{\tau_{1}^2}{\CH - {\cal D}} \right)\frac{\sqrt{g^2_{\rm
st}N}}{8} {1 \over \vert T \vert^3} \quad. \nonumber \eear
It shows that the pressure decays monotonically to that of $n$
F-strings and that the way it decays asymptotically is {\sl
independent} of the angular momentum $\ell$.

As demonstrated in the previous section, because of the angular
momentum barrier, the $(n,1)$ string bounces off if it starts at
sufficiently large distance with $\CH<\CH_{\rm cr}$. Quantum
mechanically, however, for small enough $\Delta\CH=\CH_{\rm
cr}-\CH$, the string could tunnel to the macroscopic F-string.
Reinstating the size $L_1$ of the compactified ${\mathbb{S}}_1$
direction, probability is estimated as
\bear P\sim e^{-f(\Delta\CH)L_1}\quad, \nonumber \eear
for a function $f$ determinable from the effective potentials.
This will be exponentially suppressed for large $L_1$.

%%%%%%%%%%%%%%%%%%%%%%%%%%%%%%%%%%%%%%%%%%%%%%%%%%%%%%%%%%%%%%%%%%%%%%%%%%%%%%%%%%%%%%%%
\section{Inhomogeneous Radion Rolling}
%%%%%%%%%%%%%%%%%%%%%%%%%%%%%%%%%%%%%%%%%%%%%%%%%%%%%%%%%%%%%%%%%%%%%%%%%%%%%%%%%%%%%%%%
So far, we dealt with homogeneous motion of the $(n,1)$-string. In
this section, we examine how modulation of the $(n,1)$-string
affects the dynamics. By modulation, we mean that shape of the
$(n,1)$ string in the transverse directions is initially
inhomogeneous. The effect turns out quite interesting. As the
$(n,1)$-string approaches the background F-strings, the $n$ F-string
constituents become static and the D-string constituent moves freely
in the background of $(N+n)$ F-strings. The modulation effect is
carried solely by the D-string constituent \cite{reyyee2}.

In case the worldsheet fields ${\bf X}$ being functions of both
$t$ and $\sigma$, the DBI action of the $(n,1)$-string was given
in Eq.(\ref{dbifullaction}). Explicitly,
\bear S_{\rm DBI}&=&-\tau_{1} \int \d t \d \sigma \,\sqrt{H}
\sqrt{\Big(H^{-1}-\dot{\bf X}^2\Big) \Big( H^{-1}+{\bf X'}^2\Big)
+\Big( \dot{\bf X}\cdot {\bf X}' \Big)^2-\Big((1 - H^{-1}) - {\cal
E} \Big)^2}\nonumber\quad, \eear
subject to the Gauss' law constraint $\partial_\sigma (\delta
S_{\rm DBI} / \delta {\cal E}) = 0$.

Denote the Lagrangian density as $-\tau_1{\cal L}_{\rm DBI}$.
Then, the canonical momenta are given by
\bear {\bf \Pi} &=& + \tau_{1} {1 \over \CL_{\rm DBI}} \Big[
\dot{\bf X} + H \Big( \dot{\bf X} ({\bf X}' \cdot {\bf X}') - {\bf
X}'
(\dot{\bf X}  \cdot  {\bf X}'\Big) \Big] \nonumber \\
{\cal D} &=& - \tau_{1} { 1 \over \CL_{\rm DBI}} \Big[ (H-1) - H
\, {\cal E} \Big]. \label{canonical}\eear
From the equation of motion for gauge fields and the Gauss' law
constraint, it follows that the displacement field $\cal D$ is
constant. On the other hand, the gauge potential ${\cal A}_1$ and
hence the electric field ${\cal E} = \dot {\cal A}_1$ may vary over
the worldsheet as functions of both $t$ and $\sigma$.

From the canonical momenta Eq.(\ref{canonical}), the following
relations are easily obtainable:
\bear \dot{\bf X} \! \cdot \! {\bf X}' &=& \CL_{\rm DBI}
{{\bf \Pi} \cdot {\bf X}' \over \tau_1} \nonumber \\
\dot{\bf X}^2 &=& \left(\frac{\CL_{\rm DBI}}{\tau_1}\right)^2
\left({{\bf \Pi} + H ({\bf \Pi} \cdot {\bf X}') {\bf X}'\over 1 +
H {\bf X'}^2}\right)^2. \nonumber \eear
These relations enables to express the DBI Lagrangian density
${\cal L}_{\rm DBI}$ in a compact form
\bear \CL_{\rm DBI} = {\tau_1\Big(1 + H {\bf X'}^2\Big) \over
{\sqrt{H}\sqrt{ \Big(\tau_1^2+{1 \over H} {\cal D}^2\Big)\Big(1 +
H {\bf X'}^2\Big) + {\bf \Pi}^{\rm t} {\cal M} {\bf \Pi} }}}
\quad, \label{lagcompact} \eear
where ${\cal M}$ denotes a $(8\times 8)$ matrix of dyad form
\bear {\cal M}_{ab} := \delta_{ab} + H {\bf X}'_a {\bf X}'_b.
\label{M} \eear
The Hamiltonian density ${\cal H}$ of modulated $(n,1)$-string:
\be \CH\,\,=\,\, {\bf \Pi} \cdot \dot{\bf X} + D \dot A_1
-\CL_{\rm DBI} \ee
is then easily derivable. After straightforward computation and
using the expression Eq.(\ref{lagcompact}) for ${\cal L}_{\rm
DBI}$, we obtain
\bear \CH &=& {1 \over H} \left( +(H - 1) {\cal D}+ \sqrt{ H
\Big(\big( \tau_{1}^2+\frac{{\cal D}^2}{H}\big) (1 + H {\bf X}'^2)
+ {\bf \Pi}^{\rm t} {\cal M} {\bf \Pi} \Big)  } \, \right)\quad.
\eear

Compared to homogeneous rolling of the radion analyzed in the
previous sections, a difference is that the Hamiltonian now depend
on the momenta $\Pi$ through a quadratic form defined by the matrix
Eq.(\ref{M}). In fact, the matrix Eq.(\ref{M}) admits a simple
physical interpretation. Being modulated, the $(n,1)$ string is no
longer straight along the $x^1$-direction but sweeps a contour bent
over to ${\bf X}$-directions. Thus, an infinitesimal element of the
string contour sweeps ($\Delta X^1, \Delta {\bf X})$, whose proper
length is given by
\bear \sqrt{ \Big({1 \over \sqrt{H}} \Delta X^1\Big)^2 + \Delta
{\bf X}^2 } \nonumber \eear
The consideration results in {\sl increase} of the Hamiltonian
density and hence the string energy per unit $\Delta X_1$, and {\sl
decrease} of the pressure per unit $\Delta X_1$. Intuitively, this
is explainable by the following considerations. First, the
modulation has the effect of compressing in a given $\Delta X_1$
interval more string bits, and this explains increase of the string
mass density. Second, adjacent string bits are continuously
disoriented such that tension of each bit, which is a vectorial
quantity, adds up but with slight cancellations, and this explains
decrease of the pressure. A phenomenon similar to this was also
considered in the contexts of $(p,q)$ string junction \cite{reyyee2}
and of tachyon rolling \cite{yeeyi}.

As $(n,1)$ string rolls down toward the horizon, its worldsheet
dynamics is governed by
\bear \CH_{(n,1)}\,\,\rightarrow\,\, {\cal D} + \CH_{\rm D1},
\nonumber \eear
where
\bear \CH_{\rm D1} = \sqrt{\tau_{1}^2 \Big( {1 \over H} + {\bf
X'}^2\Big)+ {1 \over H} {\bf \Pi}^{\rm t} {\cal M} {\bf \Pi}
}\quad. \nonumber \eear
The interpretation of the first term is clear; it is the tension of
the initially dissolved F1 strings. It adds to the tension $N$ of
the background F-strings to yield the tension of the $(N+n,1)$ bound
state in large $N$ limit. Observe that the second part now is
independent of any effects of gauge flux; the Hamiltonian is {\it
decoupled} into one describing $n$ F1 strings and the other
describing a D-string. We interpret this as a decoupling of F-string
constituents out of D-string in $(n,1)$ string world volume dynamics
in large $n$ limit. In other words, we propose that the world volume
theory of $(n,1)$ string in large $n$ limit takes an approximately
decoupled form,
\bear \CH_{(n,1)}\,\,=\,\,\CH_{\rm F} \,+ \,\sqrt{\tau_{1}^2\Big({1
\over H} + {\bf X'}^2\Big)+ {1 \over H} {\bf \Pi}^2 + \Big({\bf \Pi}
\cdot{\bf X}' \Big)^2}\,+\,\CO\Big(\frac{1}{n}\Big)\quad,
\label{limitH} \eear
where ${\bf X}^i$, ${\bf \Pi}_i$ are conjugate variables describing
oscillations of the D-string inside the $(n,1)$ bound-state, and
$\CH_{\rm F}$ encodes the energy of $n$ F-strings.

By Legendre transformation, we can easily get the effective action
of the decoupled D1 inside $(n,1)$ string,
\bear S_{\rm eff}\,\,=\,\,-\tau_{1}\int \d t \d \sigma \, \sqrt{{1
\over H} -\dot{\bf X}^2+{\bf X}'^2 + H \Big( (\dot{\bf X} \cdot
{\bf X}')^2 - \dot{\bf X}^2 {\bf X}^{'2} \Big) } \quad. \nonumber
\eear
Now, as the string approach the horizon, $H \rightarrow \infty$
and the last term inside the square-root forces the string
modulation to evolve in a way that $\dot{\bf X}$ is proportional
to ${\bf X}'$, and in turn that ${\bf \Pi}$ is oriented parallel
to ${\bf X}'$. This implies that the asymptotic Hamiltonian
density of the D-string constituent is governed by
\bear {\cal H}_{\rm D} =  \sqrt{\Big( {1 \over H} + {\bf
X}^{'2}\Big) } \sqrt{ \Big( \tau_{1}^2 + {\bf \Pi}^2\Big)} \,\,
\gg \,\, {1 \over \sqrt{H}} \sqrt{ \Big(\tau_{1}^2 + {\bf
\Pi}^2\Big)} \, . \nonumber \eear
The first factor is the aforementioned geometric Jacobian of the
string contour.
Evidently, the Hamiltonian density is enhanced by
this geometric factor due to the presence of the modulation, the
enhancement ratio being $\sqrt{1 + {\bf X}'^{2} /H}$.

Finally we like to mention that, at the first stage of the rolling,
the modulation effect  grows in general. This is due to typical
tidal effect in the supergravity background as is clear clear from
the static radion potential in Fig. 1.

\section{Exceptional D-Particles and Supertubes}
So far, we have dealt with dynamics of $(n,1)$ string in Type IIB
string theory. Essentially the same consideration would be
applicable for dynamics of other D-brane or bound-state of D-branes
and NS-branes in Type IIA string theory. There is, however, an
exceptional situation --- unlike all other D-branes, D-particle does
not give rise to radion rolling dynamics. This is because the
complex of D-particles and F-strings is supersymmetric for arbitrary
transverse separations, and the radion dynamics yields a motion
along flat direction. Moreover, one can manufacture a supertube
extended parallel to the macroscopic F-string. Again, such supertube
is supersymmetric in the macroscopic F-string background and hence
do not develop radion potential. In all these cases, the brane
complex preserves one-quarter of the total 32 supersymmetries. In
this section, we shall study features of these exceptional cases.
Instead of directly verifying the vanishing radion potential, we
shall demonstrate that a supertube\cite{Mateos, Lee, reyunpub} near
the macroscopic F-strings does not exhibit tachyonic instabilities.
D-particle is a special situation in which supertube loses F-string
charge.

Supertube is a bound state of D-particles, F-strings and D2-branes
in Type IIA string theory. Its worldvolume is homogeneously extended
in its axial direction along which the bound F-strings are
stretched. Interestingly the cross sectional shape of the supertubes
may form an arbitrary curve in the transverse eight dimensions
\cite{Karch}. Here, we would like to show that there is no tachyonic
instabilities for the supertube whose axial direction is aligned
with the stretching direction $X_1$ of the background fundamental
strings.

The low-energy dynamics of a supertube is conveniently described
by the DBI action of the D2 brane with electric and magnetic gauge
field excitations,
\bear S_{\rm DBI} =-\tau_{2} \int \d^3\sigma
\,e^{-(\phi-\phi_\infty)} \sqrt{-\det(X^*(G+B)+ {\cal F})}\, .
\label{tubeaction} \eear

We shall proceed with partial gauge-fixing of the worldvolume
reparametrization invariance by setting $\sigma_0=t=X^0$ and
$\sigma_1=X^1$ while leaving the $\sigma_2=\theta$ worldvolume
direction may be fixed after solving equation of motion. Consider
homogeneous dynamics along the longitudinal $\sigma^1$-direction.
Turn on the worldvolume gauge field as
\bear {\cal F}=\left[ {\cal E}(\theta, t) - \Big(1 - {1 \over H}
\Big) \right]\, \d t \wedge \d \sigma_1 + {\cal B}(\theta,t)\, \d
\sigma_1 \wedge \d \theta\, \nonumber \eear
where specific parametrization of the electric field component is
motivated by the effect of the supergravity background and
simplification in describing supertube worldvolume excitations.
Denoting the derivative $\partial_\theta$ as ${}'$, the action
density is then reduced to $-\tau_2 {\cal L}_{\rm DBI}$, where
\bear {\cal L}_{\rm DBI} = \sqrt{ \Big({1\over H}-\dot{\bf X}^2
\Big)\Big({\bf X'}^2+ H {\cal B}^2\Big) +(\dot{\bf X} \cdot {\bf
X'})^2 -H {\cal E}^2 {\bf X'}^2 -2 H {\cal E}{\cal B} \, \dot{\bf
X} \cdot {\bf X'} } \, . \eear

Again, we work with the Hamiltonian formulation. Canonical conjugate
momenta are
\bear {\bf \Pi} &=& {\tau_2 \over {\cal L}_{\rm DBI}} \Big[
\dot{\bf X} \Big({\bf X'}^2+ H\, {\cal B}^2\Big) - {\bf X}'
\Big(\dot{\bf X} \cdot {\bf X}' -  H\, {\cal E} {\cal B}\Big)
\Big]
\nonumber \\
{\cal D} &=& {\tau_2 \over {\cal L}_{\rm DBI}} H \Big( {\cal
E}\,{\bf X}'+ {\cal B} \,  \dot{\bf X} \Big) \cdot {\bf X}'
 \, .
\eear
The Hamiltonian density is then given by
\bear {\cal H} = {\cal D} \Big(1-{1 \over H} \Big)
 + \sqrt{ \tau^2_2 \Big( {1 \over H} {\bf X'}^2 +{\cal B}^2\Big) + {1 \over H} {\bf \Pi}^2 +  {1\over
H^{2}} {\cal D}^2 }\,. \nonumber \eear
We now rearrange terms inside the square-root into the following
form \cite{Ohta}:
\bear {\cal H}
={\cal D} \Big(1-{1\over H}\Big)
+\sqrt{ \bigg( \frac{\cal D}{H} + \tau_2 {\cal B}\bigg)^2 +{1\over
H } \bigg( \tau_2 \left|{\bf X}'\right| - \frac{\cal B D}
{\left|{\bf X}' \right|} \bigg)^2 + {1\over H} \bigg({\bf \Pi}^2 -
\frac{{\cal B}^2 {\cal D}^2}{{\bf X'}^2} \bigg) }\,. \label{h1}
\eear
The second term inside the square root is manifestly positive
definite. The third term inside the square root turns out positive
definite because of the Schwarz inequality:
\bear {\bf \Pi}^2 - \frac{{\cal B}^2 {\cal D}^2}{ {\bf X'}^2} =
\frac{\tau_2^2}{{\cal L}_{\rm DBI}^2} \frac{ \Big({\bf X'}^2 + H
{\cal B}^2\Big)^2}{{\bf X'}^2} \Big( \dot{\bf X}^2{\bf X'}^2 -
(\dot{\bf X} \cdot {\bf X'})^2 \Big) \geq 0\,. \eear
The equality holds when $\dot{\bf X}$ is proportional to ${\bf X}'$,
which is equivalent to $\dot{\bf X} = 0$ by a suitable worldvolume
coordinate transformation of $\theta$. We see then that the
Hamiltonian density is bounded from below and saturated when both
the second and the third terms in the square root of Eq.(\ref{h1})
vanish:
\bear
\tau_2 {\bf X'}^2 = {\cal B D}
\qquad \mbox{and} \qquad
\dot{\bf X}=0\,.
\label{BPS1}\eear
Therefore, we conclude that
\be {\cal H} \ \ge\  {\cal D} + \tau_2 {\cal B}\,. \label{ineq}
\ee
We see that the minimum of the Hamiltonian density, as given by
the right-hand side, admits a clear physical interpretation. It
contains mass density of the fundamental strings and that of the
D0-branes and precisely matches with the energy density of the
supertube! Recall that $\cal D$ is the lineal density of
fundamental strings stretched along $X^1$ direction with respect
to $\theta$ and ${\cal B}/(2\pi)$ corresponds to the area density
of D-particle inside D2-brane worldvolume. Namely, $Q_1= \int
\d\theta \, {\cal D}$ counts the number of fundamental strings and
${1\over 2\pi }\int \d\sigma^1 \d\theta\, {\cal B}$ is the number
of D-particles melted in the D2-brane.

The conditions Eq.(\ref{BPS1}) are nothing but the BPS conditions
for 1/4-supersymmetric supertubes ought to obey. Of course, one may
explicitly verify that any solution of the above BPS equations
satisfy the full equations of motion provided the Gauss law
constraint $\partial_{1} D=0$. The construction of all BPS solutions
is simple. We choose  arbitrary static functions of ${\bf
X}(\theta)$ and $B(\theta)$. Then $D(\theta)$  is constrained by
Eq.(\ref{BPS1}). Hence, the cross sectional curve is arbitrary in
the transverse eight dimensions and the distribution of D0 over the
$\theta$ direction is arbitrary too.

This demonstrates that the supertube may be located anywhere in the
eight-dimensional transversal space and there is no tachyonic
instability. From the consideration of partial supersymmetries
preserved, this fact is quite expected as discussed before, but the
actual demonstration that the BPS bound is actually saturated is
highly nontrivial as we have seen.

\vskip 1cm \centerline{\large \bf Acknowledgement} \vskip 0.5cm
S.-J.R was supported in part by the MOE BK-21 Project through SNU
Physics Division Team 2, by the MOST-KOSEF Leading Scientist Grant,
and by the Alexander von Humboldt Foundation through the F. Bessel
Research Award. D.B. is supported in part by Korea Research
Foundation Grant KRF-2003-070-C00011. H.-U.Y. is partly supported by
grant No.R01-2003-000-10391-0 from the Basic Research Program of the
Korea Science \& Engineering Foundation.

%%%%%%%%%%%%%%%%%%%%%%%%%%%%%%%%%%%%%%%%%%%%%%%%%%%%%%%%%%%%%%%%%%%

\end{document}